\documentclass[aps,pre,preprint,onecolumn,showpacs,superscriptaddress]{revtex4-1}
\usepackage{times}
\usepackage{amsmath}
\usepackage{natbib}
\usepackage{graphicx}
\usepackage{color}

\begin{document}

\title{Cooperation in public goods games: stay, but not for too long}

\author{Lucas Wardil}
\email[E-mail me at: ]{wardil@fisica.ufmg.br}
\affiliation{Departamento de F\'isica, Universidade Federal de Minas Gerais - Caixa Postal 702, CEP 30161-970, Belo Horizonte, MG, Brazil.}

\author{Marco A. Amaral}
\affiliation{Departamento de F\'\i sica, Universidade Federal do Rio Grande do Sul, Caixa Postal 15051, CEP 91501-970, Porto Alegre - RS, Brazil}
\date{\today}
\begin{abstract}
Cooperation in repeated public goods game is hardly achieved, unless contingent behavior is present.  Surely, if  mechanisms promoting positive assortment between cooperators are present, then cooperators may beat defectors, because cooperators would collect greater payoffs. In the context of evolutionary game theory,  individuals that always cooperate cannot win the competition against  defectors in well-mixed populations. Here, we study the evolution of a population where fitness is obtained in repeated public goods games and  players have a fixed probability of playing the next round. As a result, the group size decreases during the game. The population is well-mixed and there are only two available strategies: always cooperate (ALLC) or always defect (ALLD). Through numerical calculation and analytical approximations we show that cooperation can emerge if the players stay playing the game, but not for too long. The essential mechanism is the interaction between the transition from strong to weak altruism, as the group size decreases, and the existence of an upper limit to the number of rounds representing limited time availability.
\end{abstract}
\maketitle







\section{Introduction}
Cooperation in humans is  different from other animals \cite{r1}. Humans have developed cooperative strategies that rely on advanced cognitive capacities \cite{r2,r3,wardil}. The emergence of human language gave rise to  cultural evolution \cite{lg1,lg2}, with humans following norms and internalizing them as emotions. The fear of being retaliated, or being rewarded, in future interactions make individuals more cautions, or more prone to cooperate \cite{rw1,rw2}. Repetitive interactions opened new paths for the evolution of cooperation in populations. In repeated games, individuals can condition their current behavior on previous interactions. The winning strategy in the Axelrod tournament \cite{r8}, called tit-for-tat strategy (TFT starts with cooperation in the first round and does whatever the opponent did in the previous round) outcompeted all opponents in that tournament. In the context of evolutionary game theory, TFT can resist invasion by ALLD (to defect in all rounds) if the average number of rounds is not too small \cite{nowak-book}. In other words, cooperation is stable if players stay playing the game.  

The possibility of terminating an interaction and leaving the game is a feature of mobile organisms \cite{mobile}. Beneficial interactions should last and unfavourable ones should be interrupted.  In Axelrod tournaments,  the average number of rounds was determined by a constant, behavior-independent probability of playing the next round. The probability of leaving the game may depend on the behavior of the players. In a two-person repeated prisoner's dilemma game, it was shown that TFT can invade a population of ALLD if a player decision to leave the game depends on whether she cooperated or defected in the previous round  \cite{Feldman1}. Most interestingly, even ALLC can be stable against invasion of ALLD if the players' probability of continuation  after the  player herself has cooperated is sufficiently high \cite{Feldman1}.  The introduction of a continuation probability enlarges the strategy space: depending on the outcome of the current round,  should one cooperate, defect or leave the game in the next round?  With this larger set of strategies, it was shown that conditional dissociation allows the coexistence of defectors and a type of strategy that responds defection by leaving the game \cite{oft}. Several other models have been proposed to study conditional leaving, but they all combine different mechanisms, making it hard to isolate the effect of conditional dissociation \cite{hayashi,aktipis,hamilton}. We must stress that the possibility of terminating an interaction is not always possible, for example, when population has static spatial structure resulting in a fixed neighborhood. In this case, spatial structure may create reciprocity, favouring cooperation \cite{sp1,sp2,sp3,sp4}. 

Instead of leaving an ongoing game,  individuals may have the option to stay aside, not playing the game, and rely on a small but fixed income. Research on  exit of public goods games usually focus on how outside options affect contributions.  Both positive and negative consequences of exiting having been pointed out \cite{optout1,optout2,optout3,optout4,optout5,optout6}. A common feature of these models is that player can join the game after exiting.  Interestingly, most models study collective behaviour in groups of size two. One reason for this choice is that there is no complete understanding of the effect of the group size on public goods games  yet. Terminating a pair-wise interaction implies the end of the game, but opting-out  when there are more players  raises the question of how to define the public goods benefits. One possibility is to rescale the multiplicative factor in order to keep the marginal return constant \cite{gsize,optout1}. In this case, the social dilemma is present independently of the group size.

Another approach to deal with variable group size is to keep the multiplicative factor constant. In this case, the marginal return can vary and the game may change from a regime where defection is the best option to a regime where cooperation is best one. More specifically, in a public goods game, a cooperator contributes $1$ token to  a common endeavour. The total contribution is multiplied by $r$ and is evenly split among the $M$ members of the group. Thus, cooperators receive $rk/M$ and defectors receive $rk/M+1$, where $k$ is the number of cooperators in the group. Each one unit contributed to the pool returns $r/M$ units to the contributor. Therefore, if $M<r$, then the best choice is to contribute. The regime $M<r$ is called weak altruism \cite{weakalt}. It is a weak altruism because it is of the best interest for cooperators to contribute and, at the same time, defectors do better at the expenses of  cooperators. This possibility was explored in an experiment with humans and, theoretically, in the context of evolutionary dynamics. If cooperators, defectors, punishers and nonparticipants are the available types, it was shown that  nonparticipants pave the way for cooperation in the context of public goods game \cite{freedom,volunt-exp}. In the absence of punishers, non-participants spread in the population at first. Since most players do not participate in the game, the average size of the groups playing the public goods game is small and the social dilemma disappears: the payoff of cooperators are, on average, larger than the payoff of defectors. The advantage for cooperators appears when groups of size $M$ are randomly sampled in a population. In small groups cooperators take advantage of the weak altruism regime and, on average, do better than defectors \cite{pgg}.
  
In repeated public goods game, if  players leave the group during the game, then the group size may decrease and the weak altruism threshold may be reached. If players do not return to the game, the weak altruism regime may last for a sufficiently large number of rounds, conferring advantage to cooperators.   Here, we study the evolutionary dynamics of a well-mixed population where fitness is obtained in repeated public goods games. Individuals play the next round  with a probability $w$ that is constant and independent of the behavior and do not return to the game after leaving. The multi-round game is over either because the group vanishes or because a maximum number of rounds is reached, which is set to constant value. There are only two unconditional strategies: ALLC and ALLD.  We show that cooperation is favored  by selection if players stay playing the game, but not for too long. 

\section{The model}
 
The game consists of several rounds of a public goods game. Initially, $M$  players are randomly selected in a well-mixed population of size $N$ to play the multi-round public goods game. Each player plays the next round with probability equal to $w$. So, the group size can decrease throughout the rounds. If the number of players is less than two, the game is over. The maximum number of rounds is set to a fixed number, $n_{max}$, to avoid the case of infinite repeated games if $w=1$. There are two strategies:  cooperation ($C$) and  defection ($D$). At the beginning of each round,  each player that is participating in the game has an initial endowment of one token. Cooperators contribute with one token to a common pool. The total of contributions is multiplied by $r$ and is  split evenly among all the participants, including defectors. Defectors keep the initial endowment. Let $M_n$ be the number of individuals playing the game in round $n$ and $k_n$ be the number of cooperators in this group. The payoffs of the cooperators and the defectors that are playing the game in  round $n$ are given by
\begin{eqnarray}
\nonumber
   \pi_C^n&=&\frac{rk_n}{M_n}, \label{pay1} \\ 
   \pi_D^n&=& \frac{rk_n}{M_n}+1. \label{pay2}
   \label{payoff_singleround}
\end{eqnarray}
Players accumulate their gains through the rounds and stop to collect the total payoff when they leave the game.  Following the usual procedure of evolutionary game dynamics, the cumulative payoff is interpreted as reproductive success and a Moran process is implemented \cite{nowak-book}. At the end of each multi-round game, an individual is randomly chosen to die and another individual is chosen to reproduce with a probability proportional to $1-s+s\pi$, where $\pi$ is the cumulative payoff and $s$ is the intensity of selection. After the reproduction step, the payoff of each player is equated to zero and, again, $M$ players are randomly selected to play a new multi-round game. Players that are not selected to play the multi-round game have cumulative payoff equal to zero.  Notice that the payoff defined in Equations \ref{payoff_singleround} comply with the usual definition adopted in experiments with humans, differing from the alternative definition where, instead of an initial endowment, there is a cost to contribute to public good \cite{freedom}. The results in our work are the same for both payoff definitions.

This model sets up the most severe scenario for cooperation in terms of the relation between interaction rate and selection rate. It was shown that, if only  few individuals collect payoff before a reproduction step takes place (recall that, in our model, a game is actually a multi-round one), then the fixation probability  of a cooperator  is drastically reduced. In contrast, if all players have the opportunity to play the game, then cooperation has a higher chance to survive \cite{roca}. Since in our model only $M$ individuals  play the multi-round game before a reproduction step takes place, we are at the most severe scenario. We also checked the model where the process of selecting $M$ players is repeated several times before the reproduction step. Since it does change our results qualitatively, we do not present the analysis of this variation here. 

\section{Results}

To see if natural selection  favors cooperation in finite well-mixed populations, we  look at the probability that a single cooperator in a population of defectors turns the population to full cooperation. This probability is called  fixation probability \cite{nowak-book}, and is given by
\begin{equation} \label{eq1}
\rho=\frac{1}{1+\sum_{k=1}^{N-1}\prod_{n_c=1}^k \frac{1-s+s\left<\pi_{D}(n_c)\right>}{1-s+s\left<\pi_{C}(n_c)\right>}},
\end{equation}
where $\left<\pi_{C}(n_c)\right>$ and $\left<\pi_{D}(n_c)\right>$  are the average cumulative payoff of a cooperator and a defector, respectively, in a population of size $N$ with $n_c$ cooperators. If the game plays no role in the evolution of strategies, that is, if $s=0$, then the fixation of a mutant is driven only by random drift and is given by  $\rho=1/N$. If $s\ne 0$, we say that cooperation is favored by selection if $\rho>1/N$ and  that cooperation is inhibited by selection if $\rho<1/N$ \cite{nowak-book}. 

\begin{figure}[h]
\vspace{1cm}
  \centering
  \centerline{\scalebox{0.4}{\includegraphics{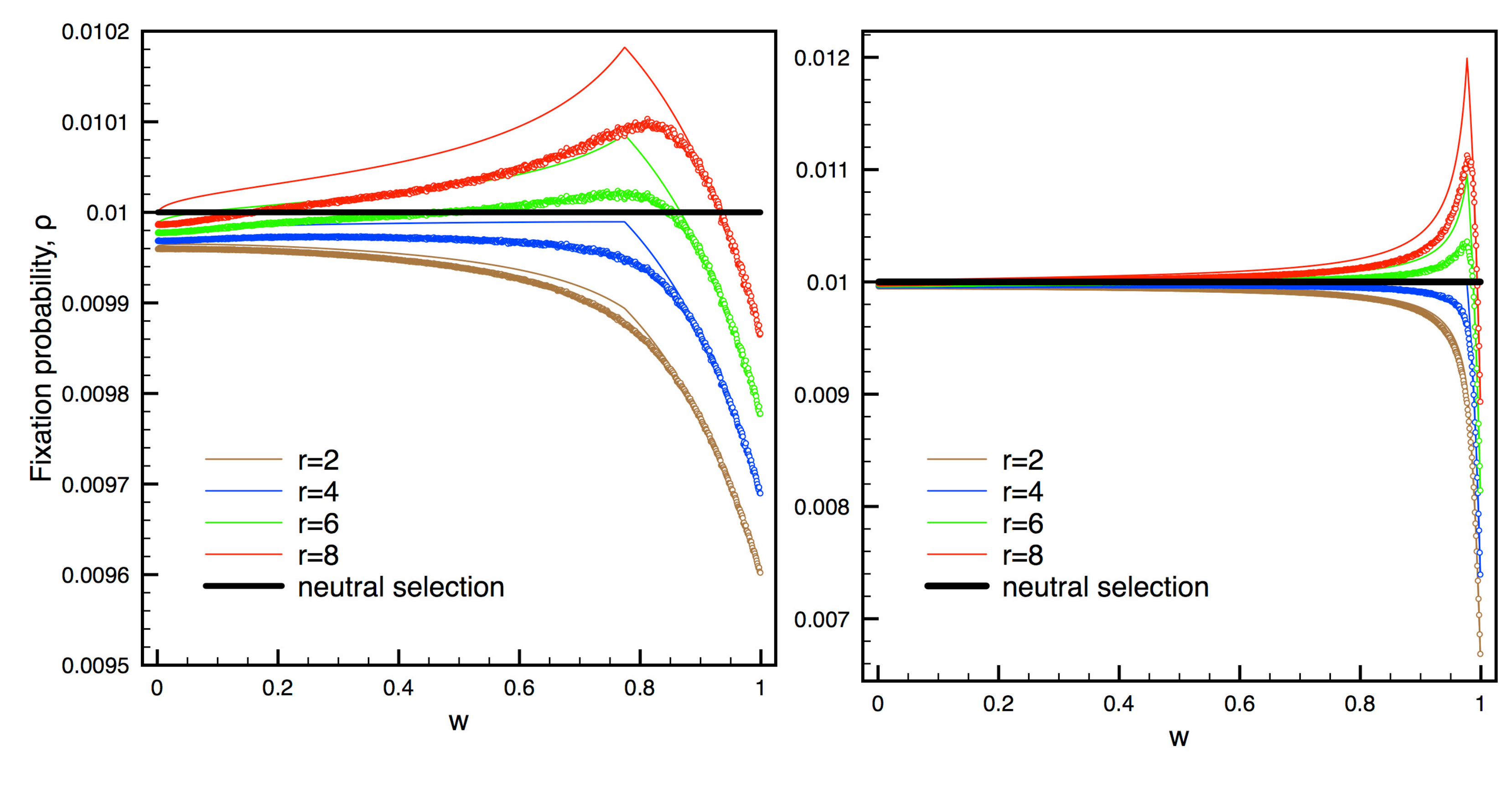}}}
\caption{Fixation probability, $\rho$, of a single cooperator in the weak selection limit for $n_{max}=10$ (left) and $n_{max}=100$ (right). We show data from computer simulation (symbols) and from the analytical approximation, Equation \ref{eqweak} (continuous lines). The horizontal line at $\rho=0.01$ is the neutral fixation probability. The  parameters are  $N=100$, $M=10$, and $s=0.001$.}\label{fig1}
\end{figure}

The weak selection limit is usually  appropriated  to incorporate the game payoffs into the fitness, since we do not expect the game to be a major fitness component. 
In the limit of weak selection, a Taylor approximation for $s \rightarrow 0$ on Equation \ref{eq1} leads to
\begin{equation}
\label{weakapprox}
\rho \approx \frac{1}{N}\left[1+s\sum_{n_c=1}^{N-1}\frac{N-n_c}{N}\left(\left<\pi_C(n_c)\right>-\left<\pi_D(n_c)\right> \right) \right].
\end{equation}
To calculate the fixation probability in the limit of weak selection using Equation \ref{eq1}, we must have the expression for  $\left<\pi_{C}(n_c)\right>$ and $\left<\pi_{D}(n_c)\right>$. Since we can not derive a closed expression for the average payoffs, we first solved Equation \ref{eq1} numerically, with the values of $\left<\pi_{C}(n_c)\right>$ and $\left<\pi_{D}(n_c)\right>$ obtained from computer simulations of the payoff accumulation phase.  Second, we developed an analytical approximation, starting with Equation \ref{weakapprox} (see below in this section). Figure \ref{fig1} shows the fixation probability as a function of $w$  in the weak selection regime for both the numerical and analytical methods.  Note that, for small $w$, the  fixation probability is quite close to the neutral fixation probability ($\rho=1/N$). For sufficiently large $r$, the  probability increases, until it reaches a maximum. After this point, the probability decreases to values lower than the neutral fixation probability, as $w$ tends to $1$. Notice that the probability of fixation and the value of $w$ at which the fixation is maximum increase for higher values of $n_{max}$  So, to increase cooperation in a population constrained to a fixed maximum number of rounds, players should stay in the game, but not for too long. 

To understand these results, let us  derive an analytical approximation for this model. In a population of $N$ players, with $n_c$ cooperators, the probability that there are $k$ cooperators in the randomly selected group of size $M$ in the first round is given by the hypergeometric distribution
\begin{equation}\label{prob}
H(k,M,n_c,N)=\frac{  \left( \begin{array}{c} n_c \\ k \\\end{array} \right)
                        \left( \begin{array}{c}N-n_c \\ M-k \\ \end{array} \right)  }
                     {  \left( \begin{array}{c} N \\  M \\ \end{array} \right)      }
.\end{equation}
Thus, the average  payoff of  cooperators and defectors  in the first round is given by 
\begin{eqnarray*}
  \pi_C^1 &=& \sum_{k=0}^M H(k,M,n_c,N)\frac{kr}{M}\frac{k}{n_c}, \\
  \pi_D^1 &=& \sum_{k=0}^M H(k,M,n_c,N)\left(\frac{kr}{M}+1\right)\frac{M-k}{N-n_c} 
\end{eqnarray*}
The term $\frac{k}{n_c}$  stands for the fraction of cooperators playing the game. This term guarantees the null payoff of players that are not selected to play the game. The same holds for the term $\frac{M-k}{N-n_c}$, which is the fraction of defectors playing the game. We stress that, in other models of public games in finite population \cite{freedom}, all players have the opportunity to play the game. Therefore, the average payoff of a cooperator, for example, is calculated supposing that one of the players in the group is already a cooperator, implying that $ H(k,M-1,n_c-1,N-1)$ should be used. Note that this is not our case. Returning to our model, a simple calculation leads to
\begin{eqnarray}
\label{eqpiC1}
  \pi_C^1 &=&  \frac{r}{N}\left[1+(n_c-1)\frac{M-1}{N-1}\right],\\
   \label{eqpiD1}
  \pi_D^1&=&\frac{M}{N-n_c}\left[1+(r-1)\frac{n_c}{N}\right]-\frac{n_c}{N-nc}\pi_C^1.
\end{eqnarray}
The average group size in round $n$ is given by $M_{n}=M_{n-1}w$. This  recurrence relation yields
\[
M_n=Mw^{n-1}.
\]
Now we use the approximation of replacing $M$ by $M_n$ in Equations \ref{eqpiC1} and \ref{eqpiD1}. Within this approximation, the average payoff of  cooperators and adefectors in round $n$ is given by
\begin{eqnarray*}
  \pi_C^n &=&  \frac{r}{N}\left[1+(n_c-1)\frac{Mw^{n-1}-1}{N-1}\right],\\
  \pi_D^n&=&\frac{Mw^{n-1}}{N-n_c}\left[1+(r-1)\frac{n_c}{N}\right]-\frac{n_c}{N-nc}\pi_C^n.
\end{eqnarray*}
The multi-round game is over when there is less that two players or when the maximum number of rounds $n_{max}$ is reached. The average number of rounds, $ n^*$, is given by
\begin{equation} 
\label{eqnast}
n^*=\min\left(n_{max},1+\frac{\ln(1/M)}{\ln(w)}\right) \;\;\; \mbox{ for } \;\;\; 0<w<1.
\end{equation}
If $w=0$, the average number of rounds is given by $n^*=1$ and if $w=1$ it is given by $n^*=n_{max}$. Therefore, the average cumulative payoff at the end of $n^*$ rounds is  given, within our approximation, by
\begin{eqnarray}
\label{eqaveC}
 \nonumber
  \left<\pi_{C}\right> &=& \sum_{n=1}^{n^*}\pi_C^n \\
            &=& \frac{r}{N}n^*\left(1-\frac{n_c-1}{N-1}\right)+\frac{r}{N}\frac{n_c-1}{N-1}M\frac{w^{n^*}-1}{w-1}
\end{eqnarray}
\begin{eqnarray}
\label{eqaveD}
 \nonumber
 \left< \pi_{D}\right>&=& \sum_{n=1}^{n^*}\pi_D^n \\ 
  &=& \frac{M}{N-n_c}\frac{w^{n^*}-1}{w-1}\left(1+(r-1)\frac{n_c}{N}\right)-\frac{n_c}{N-n_c}\left<\pi_{C}\right>
\end{eqnarray}

\begin{figure}[h]
\vspace{1cm}
  \centering
  \centerline{\scalebox{0.8}{\includegraphics{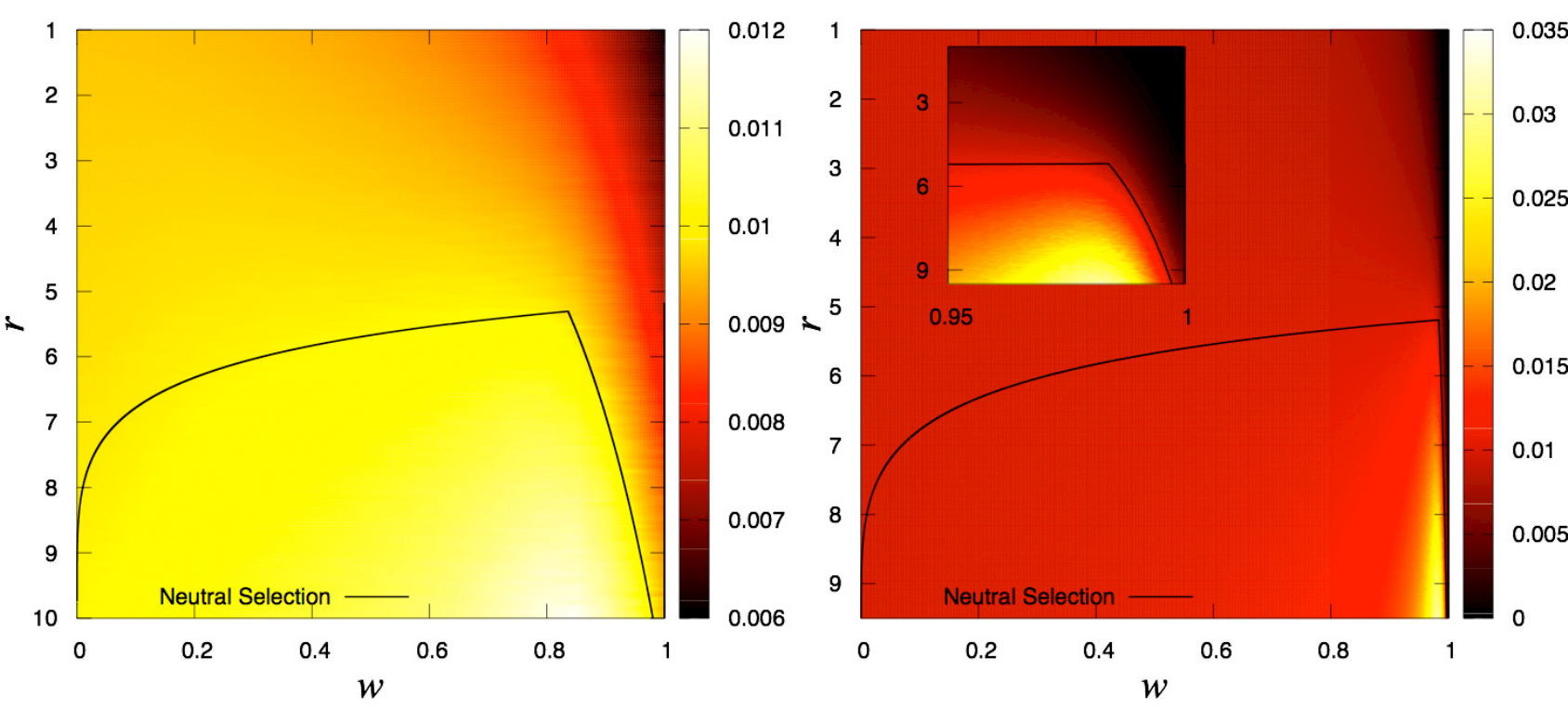}}}
\caption{Fixation probability, $\rho$, as a function of $r$ and $w$ for $n_{max}=10$ (left) and $n_{max}=100$ (right) obtained from computer simulations. The solid line shows the analytical value where the fixation probability equals neutral selection, calculated using Equation \ref{eqweak}. This analytical line divides the phase space in two regions: cooperation is favored where $\rho>1/N$ and cooperation is inhibited where $\rho<1/N$. Note that, although cooperators benefit from large $w$, there is a maximum value after which it is better to leave the game. On the right figure the inset shows details of the region near $w^{peak}$. }\label{figphase}
\end{figure}

The analytical approximation of the fixation probability in the limit of weak selection is obtained  after the insertion of Equations \ref{eqaveC} and \ref{eqaveD} in equation \ref{weakapprox}. Interestingly, the term $\left<\pi_{C}\right>-\left<\pi_{D}\right>$ does not depend on $n_c$, that is,
\[
\left<\pi_C\right>-\left<\pi_D\right>=\frac{1}{N(N-1)}\left[ M\frac{w^{n^*-1}}{w-1}(N+r-1)-n^*Nr  \right].
\]
So, Equation \ref{weakapprox}  simplifies to
\begin{equation}
\label{eqweak}
\rho \approx \frac{1}{N}\left[ 1-\frac{s}{2N}\left(M\frac{w^{n^*}-1}{w-1}(N+r-1)-n^*Nr\right) \right].
\end{equation}
Selection favours cooperation if $\rho>1/N$. This inequality is satisfied if
\begin{equation}
\label{eqphase}
M\frac{w^{n^*}-1}{w-1}(N+r-1)<n^*Nr. 
\end{equation}
Inequality \ref{eqphase} defines two regions in the parameter space $w \times r$. In one region, selection favours cooperation and in the other one selection inhibits cooperation. Figure \ref{figphase} shows the probability of fixation in the parameter space $w \times r$. Notice that transforming the inequality \ref{eqphase} into an equality yields the curve that separates both regions.

Cooperators win the competition against defectors if their cumulative payoffs are greater than the payoff of defectors. Indeed, cooperators have larger cumulative payoff if  inequality \ref{eqphase} is satisfied, which is the same inequality that defines the region where selection favors cooperation. Inequality \ref{eqphase} can be rewritten as
\[
\sum_{n=1}^{n^*}(N+r-1)M_n<\sum_{n=1}^{n^*}Nr.
\]
Since $Nr$ is a constant term and $M_n$ is monotonically decreasing (recall $M_n$ is an average quantity), there must exist a $n^{\dagger}$ such that, if $n>n^{\dagger}$, then  $(N+r-1)M_n<Nr$. That is,
\begin{equation}
\label{eqMn}
M_n<\frac{Nr}{N+r-1} \;\;\; \mbox{ for }\;\;\; n>n^{\dagger}.
\end{equation}
The value of $n^{\dagger}$ can be obtained by replacing the expression of $M_{n^{\dagger}}$ ( recall $M_{n}=Mw^{n-1}$) into inequality \ref{eqMn}:
\begin{equation}
n^{\dagger}=1+\frac{\ln{\frac{Nr}{M(N+r-1)}}}{\ln{w}}.
\end{equation}
The round at which the group size is small enough and cooperators start to do better than defectors is implicitly defined in Inequality \ref{eqMn}. The meaning of this inequality is  more clear in the limit of large $N$, where it simplifies  to 
\[M_n<r.\] 
In other words, from the moment that the group size is smaller than a critical size, cooperators do better than defectors. In the limit of large population, the critical size is equal to the multiplicative factor $r$, a result that is characteristic of  weak altruism in finite populations. Hence, cooperators should stay playing the game so they can reach  the weak altruism regime. 

\begin{figure}[h]
\vspace{1cm}
  \centering
  \centerline{\scalebox{0.5}{\includegraphics{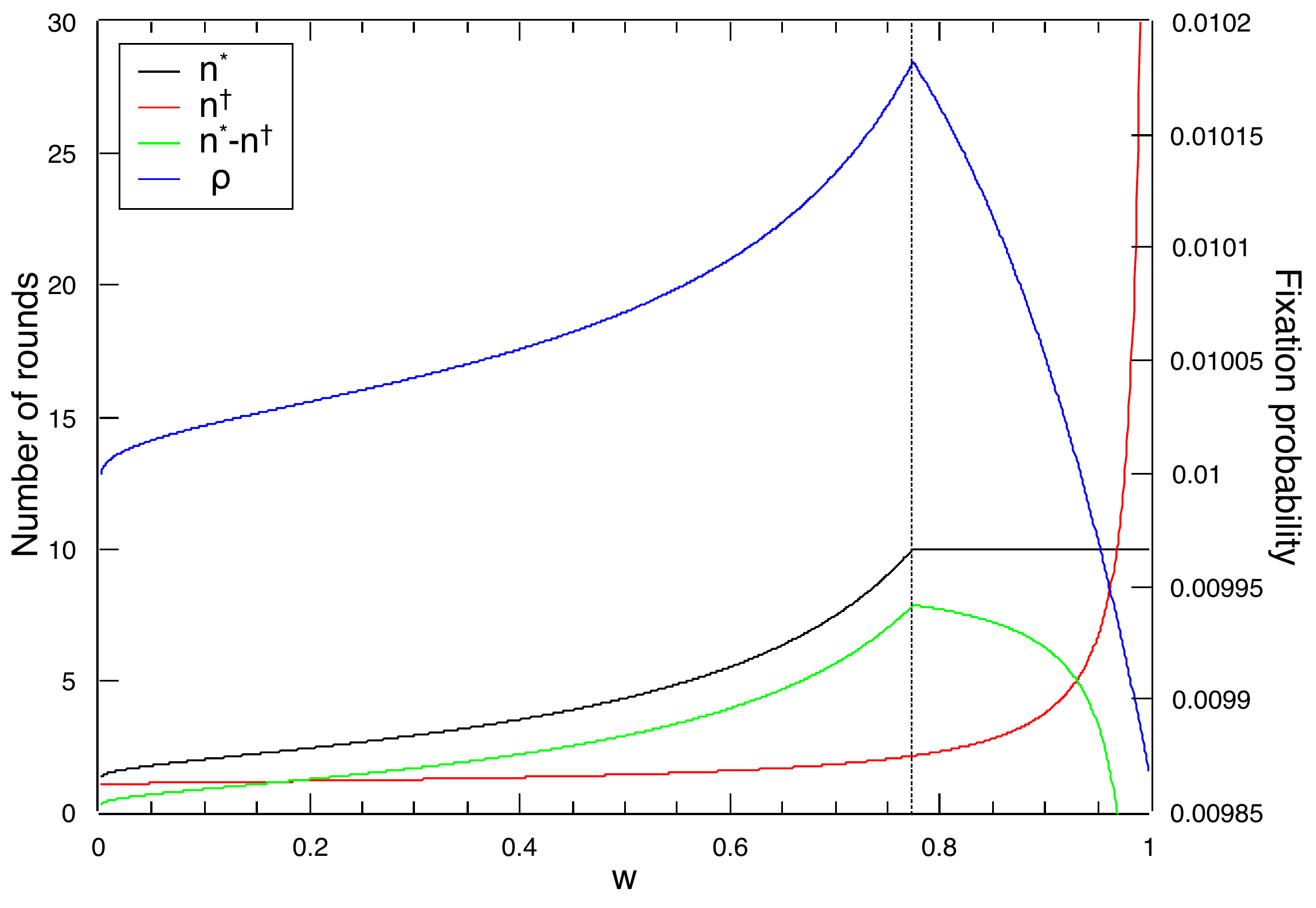}}}
\caption{Number of rounds analysis. The figure shows the average number of rounds, $n^*$; the round at which the average payoff of cooperators start to be greater than the average payoff of defectors,  $n_{\dagger}$; the difference $n^*-n^{\dagger} $; and the fixation probability for $r=8$ and $n_{max}=10$. As expected, $n^*$ grows with $w$, but reaches the constant value of $n_{max}$, which was set to $n_{max}=10$ in this figure. Notice that  the peak of $n^*-n^{\dagger}$ is at $w_{peak}$ (vertical dashed line).}\label{rodadas}
\end{figure}

Note that, to enter in the weak altruism regime, the group size must have decreased so Inequality \ref{eqMn} is satisfied before $n_{max}$ is reached. If $w$ is small, the game is over before reaching $n_{max}$. On the other hand, as $w$ approaches one, the group size remains almost unchanged, round $n_{max}$ is reached and the group size remains above the critical size defined by Inequality \ref{eqMn}. More specifically,  on average, defectors do better than  cooperators when $1\le n<n^{\dag}$ and  cooperators do better than defectors  when $n^{\dag}\le n\le n^*$.  The number of rounds where cooperators do better is given by $n^*-n^{\dagger}$.  Figure \ref{rodadas} shows the graph of $n^*$, $n^{\dagger}$,  $n^*-n^{\dagger}$ and $\rho$ as a function of $w$. Notice that $n^*$ grows faster than $n^{\dagger}$ as $w$ increases. Therefore, as $w$ increases, cooperators  play more rounds where they do better than defectors.   However, $n^*$ increases only until $n_{max}$, putting a limit on the positive effect that increasing $w$ has on cooperation. In other words, as $w$ gets closer to one, most players keep playing the game and $n_{max}$ is reached before the group size enters in the weak altruism regime.  The value of $w$ at which $n^*$ reaches $n_{max}$ is obtained from Equation \ref{eqnast}, and is given by
\begin{equation}
n_{max}=1+\frac{\ln{1/M}}{\ln{w^{peak}}}.
\label{eqnmax}
\end{equation}
After a simple transformation we have
\begin{equation}
w^{peak}=\left(\frac{1}{M}\right)^{\frac{1}{n_{max}-1}}.
\end{equation}
For $w>w^{peak}$, $n^*-n^{\dagger}$ starts to decreases and the probability of fixation becomes a decreasing function in the interval $(w^{peak},1]$. In other words, $w^{peak}$ is the value of $w$ at which the probability of fixation is maximum. Since $w^{peak}$ approaches one as $n_{max}$ goes to infinity, the values of $w$ where the probability of fixation is a decreasing function vanishes. Hence, if there is no constraint in the maximum number of rounds, cooperators should stay playing. 

\section{Discussion and Conclusion}

Repeated interactions create the possibility of leaving the game as a strategic action. If the decision of leaving is conditioned on partners behavior, there can be  assortment between cooperators and it is more likely that cooperation will be favored. In contrast, in our model the probability of leaving the game is a constant, behavior-independent parameter. In our model there is no assortment and no distinction is made between cooperators and defectors. 

The probability of continuation can be interpreted as an environmental  effect that prevents continuation in long repeated interactions. In our model, it is an external parameter upon which players have no control. If each player can decide with which probability they should play the next round, similar to what was done in \cite{oft}, the equilibrium analysis of a multi-round game implies that defectors should stay playing  all rounds. After all, why should a defector leave the game if it is always beneficial to participate in it? It is the cooperators  that should opt out if they believe it is not worth to participate in the game. We stress that this is not the case of the present study, since we focus only on non-contingent strategies subjected to equal environmental effect. 


The consequence of leaving the game is to reduce the group size. The critical group size where the dilemma changes from strong to weak altruism determines the fate of cooperation. We obtained the critical threshold as a function of $N$ and $r$. In the limit of large $N$, the threshold is simply $M_n<r$. The Nash equilibrium is to defect  if $M_n>r$ and to cooperate if $M_n<r$. In terms of absolute fitness, contribution increases an individual's payoff if $M_n<r$ (the return of one contributed token is greater than keeping it). However, in terms of relative fitness, defectors do better than cooperators in a single multi-round game since, by Equation \ref{payoff_singleround}, we have $\pi_D^n=\pi_C^n+1$ in all rounds. It is only  through the sampling process that cooperators have the chance to earn more than defectors. This is called  the Simpson's paradox \cite{simpson}, a phenomena that happens when a trend appears in isolated groups (payoffs in each single-round public goods game), but disappears when combined (the average payoff). 

The probability of continuation determines whether or not the group will reach the weak altruism regime.  If the players leave the game quickly, the group enters in the weak regime after  a few rounds. However, the group will also disappear quickly and cooperators will not enjoy the benefits of the weak altruism regime. For greater persistence, more rounds in the weak altruism regime are played and cooperators accumulate larger payoffs. The immediate consequence is an increase on the fixation probability to values greater than $1/N$. If there was an unlimited number of rounds, the advantage of cooperators would continue to increase. However, there can not be infinite rounds. If the number of rounds reaches its maximum, the group size may not have its size reduced  in time. If players start playing for too long, due to the maximum number of rounds, the fixation probability decreases and selection starts to inhibit cooperation. Therefore, to say that cooperators should not play for too long is equivalent to say that $w$ should not be greater than its critical value, with the critical value depending on $n_{max}$.

So, if players start playing for too long, due to the maximum number of rounds, the fixation probability decreases and selection starts to inhibit cooperation. We stress that mechanisms that yield differential payoff accumulation, as we studied here, are important to define the fate of cooperation in a population. For example, in \cite{javarone},  a different mechanism creates a  payoff accumulation scenario favourable to cooperation. In their mean field approach, there are two groups: in one there is only cooperators (ALLC) and in the other there are both cooperators and defectors (ALLD). If cooperators play for enough time, they accumulate large payoffs and, after a migration, they may invade the other group.

To sum up, we showed that cooperation is favored in repeated interaction if individuals stay playing, but not for too long. The basic mechanism is the interaction between the transition from strong to weak altruism and the existence of an upper limit to the number of rounds representing limited time availability.

\acknowledgments{The author  thanks to CNPq and FAPEMIG, Brazilian agencies.}







\end{document}